\documentclass[aps,pre,floatfix,showpacs,amsmath,amssymb,endnote,twocolumn,superscriptaddress]{revtex4-1}
\usepackage{graphicx}

\begin{document}

\title{Glass transition of charged particles in two-dimensional confinement}

\author{Anoosheh Yazdi}\affiliation{Institut f\"{u}r Materialphysik im Weltraum, Deutsches Zentrum f\"{u}r 
Luft- und Raumfahrt, 51170 K\"{o}ln, Germany} \affiliation{Max-Planck-Institut für extraterrestrische Physik, 
85741 Garching, Germany}

\author{Marco Heinen}
\affiliation{Division of Chemistry and Chemical Engineering, California Institute of Technology, Pasadena, California 91125, USA}

\author{Alexei Ivlev}\affiliation{Max-Planck-Institut für extraterrestrische Physik, 85741 Garching, Germany}

\author{Hartmut L\"owen}\affiliation{Institut f\"ur Theoretische Physik II: Weiche Materie, Heinrich-Heine-Universit\"at, 40225 
D\"usseldorf, Germany}

\author{Matthias Sperl}
\affiliation{Institut f\"{u}r Materialphysik im Weltraum, Deutsches Zentrum f\"{u}r Luft- und Raumfahrt, 51170 K\"{o}ln, Germany}

\date{\today}

\begin{abstract}
The glass transition of mesoscopic charged particles in two-dimensional confinement is studied by mode-coupling theory. We 
consider two types of effective interactions between the particles, corresponding to two different models for the distribution of 
surrounding ions that are integrated out in coarse-grained descriptions. In the first model, a planar monolayer of charged 
particles is immersed in an unbounded isotropic bath of ions, giving rise to an isotropically screened Debye-H\"uckel- (Yukawa-) 
type effective interaction. The second, experimentally more relevant system is a monolayer of negatively charged particles that 
levitate atop a flat horizontal electrode, as frequently encountered in laboratory experiments with complex (dusty) plasmas. A 
steady plasma current towards the electrode gives rise to an anisotropic effective interaction potential between the particles, 
with an algebraically long-ranged in-plane decay. In a comprehensive parameter scan that covers the typical range of 
experimentally accessible plasma conditions, we calculate and compare the mode-coupling predictions for the glass transition in 
both kinds of systems.
\end{abstract}


\maketitle
\section{Introduction}\label{sec:Intro}

Two-dimensional (2D) configurations of mesoscopic charged particles can be observed in various kinds of experiments 
\cite{IvlevBook}, including colloidal suspensions confined to interfaces or between plates \cite{Pieranski1983,Chang1988}, or 
negatively charged dust particles levitating in the weakly ionized plasma sheath atop and parallel to a flat horizontal electrode 
\cite{Thomas1994}. In coarse-grained descriptions one is interested in the charged particle's dynamics and phase behavior without 
taking explicit account of the surrounding electrons and ions that ensure overall charge-neutrality of the system. In this 
article, we employ mode-coupling theory (MCT) to study vitrification in two kinds of confined, monodisperse charged-particle model 
systems. The first is the traditional two-parametric model of confined particles that interact via screened Coulomb (Yukawa) 
pair-potentials, and the second is a more realistic, three-parametric model for a monolayer of negatively charged particles 
embedded in a flowing plasma.

The simple Yukawa model has been widely used in the description of dusty plasmas (see Refs.~\cite{IvlevBook,Fortov2005, 
FortovMorfillBook}). It is capable of describing the effective pair-potential between charged particles rather accurately around 
the most common (mean geometric) nearest neighbor distance \cite{Konopka2000, Kompaneets2007}. Nevertheless, the Yukawa model is 
not justified in many of the common laboratory experiments with 2D confinement, due to a highly anisotropic distribution of ions. 
In the common case of dusty plasmas, levitating in a collisional plasma sheath atop an electrode in a radio frequency chamber 
\cite{FortovMorfillBook}, account has to be taken of the plasma current of ions towards the electrode and the corresponding 
anisotropic effective dust interaction potentials. A kinetic theory of the ion distributions and effective dust grain interactions 
is appropriate in this case, and has been studied by different groups of researchers, under different assumptions on the plasma 
parameters \cite{Vladimirov1995, Vladimirov1996, Ishihara1997, Xie1999, Lemons2000, Lapenta2000, Schweigert2001, Kompaneets2007, 
Kompaneets2008}. The theory is based on the solution of the kinetic equation for ions moving in the electrostatic field of the 
sheath. Different approximations used for the ion collision operator (describing the interaction with neutral gas) merely reflect 
different experimental regimes (in terms of the radio frequency discharge power and pressure) when the particular model is 
applicable.

Among these kinetic models, the one published by Kompaneets \textit{et al.} \cite{Kompaneets2007} is based on a reasonable 
assumption of a mobility-limited ion drift in the sheath field (as opposed to rather unrealistic inertia-limited motion) and 
employs a velocity-independent ion-neutral collisional cross-section which is logarithmically accurate for the dominant 
charge-exchange collisions \cite{Fortov2005}. The resulting three-parametric potential is anisotropic in three dimensions (3D); 
for charged particles confined to 2D, it exhibits an algebraically long-ranged $r^{-3}$ decay. This model is expected to provide a 
realistic description of interactions in ground-based dusty plasma laboratory experiments \cite{Roeker2012}.

Our results, reported in the present paper, predict qualitatively similar liquid-glass transition curves for monolayers with 
Yukawa-like and Kompaneets-like pair potentials. However, we find that a glass transition in a dusty plasma monolayer may be 
qualitatively misinterpreted if Yukawa-like interactions are assumed: An apparently re-entrant liquid-glass-liquid state sequence 
is found in the parameter space of the Yukawa potentials that at distances close to the mean geometric distance best fit the 
potential derived from the kinetic theory. This apparent re-entrant state sequence is merely an artifact that arises when one 
attempts to describe the system in terms of the inappropriate Yukawa potential parameters, and it disappears when the more 
realistic kinetic potentials are assumed, and the corresponding dimensionless parameters are used in plotting the transition 
diagram.

The article is organized as follows: In Sec.~\ref{sec:Model_Systems} we discuss the two model systems of charged particle 
monolayers with Yukawa and Kompaneets pair potentials. Section~\ref{sec:MCT} provides a brief summary of the MCT equations and 
their only input, the 2D static structure factors, which are computed in the approximate T/2-HNC scheme. Our results are presented 
in Sec.~\ref{sec:Results}, preceding our finalizing conclusions in Sec.~\ref{sec:Conclusions}.

\section{The two model systems}\label{sec:Model_Systems}

Both model systems that are described in the following two subsections contain mesoscopic charged particles confined to a 2D 
plane. The charged particles' diameter is in the order of microns. Surrounding ions are only implicitly accounted for, through 
their influence on the effective pair-potential between the confined, charged particles. In the thermodynamic limit, both the 
number, $N$, of particles and the area, $L^2$, of the confining plane diverge to infinity at a fixed value of the areal particle 
number density $n = N/L^2$.

\subsection{Yukawa monolayer}

\begin{figure}[hbt]
\includegraphics[width=0.95\columnwidth]{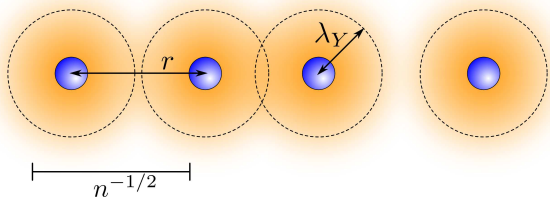}
\caption{\label{fig:Yukawa_schematic}Edge-on schematic of a Yukawa monolayer. Charged particles (filled circles) are confined to a 
plane, while oppositely charged ions are free to move in the surrounding, unbounded 3D space. The mean ion density is color-coded. 
Typical in-plane nearest neighbor distances are similar to the mean geometric distance $n^{-1/2}$, and of the same order of 
magnitude as the Yukawa screening length $\lambda_Y$. Particle separations greatly exceed the particle diameter. The effective 
particle interactions are quantified by the two dimensionless parameters $\Gamma_Y$ and $\kappa_{ Y} = 1 / (\lambda_{Y} 
\sqrt{n})$.}
\end{figure}
The Yukawa monolayer model implicitly assumes thermodynamic equilibrium statistics of ions, as schematically depicted in 
Fig.~\ref{fig:Yukawa_schematic}. Unlike the two-dimensionally confined, mesoscopic charged particles, ions are free to move in 3D 
space in the absence of external forces. Under these conditions the effective interaction potential $U_Y(x)$ between charged 
particles at sufficiently large mutual distance follows the screened Coulomb (Yukawa)-type form \cite{Khrapak2010}
\begin{equation}\label{eq:Y_pot}
 \frac{U_Y(x)}{k_\text{B}T} = \Gamma_{Y} \dfrac{\exp(-\kappa_{Y} x)}{x},
\end{equation}
where $k_B$ is Boltzmann's constant, $T$ is the absolute temperature, and $x = r \sqrt{n}$ is the particle center-to-center 
distance in units of the mean geometric distance $n^{-1/2}$.

The Yukawa potential in Eq.~\eqref{eq:Y_pot} is characterized by the two dimensionless parameters $\Gamma_Y$ and $\kappa_Y$: The 
coupling parameter $\Gamma_Y = Q_Y^2 \sqrt{n} / (4 \pi \epsilon k_\text{B} T)$ quantifies the interaction strength in terms of the 
charged particle's effective Yukawa charge $Q_Y$ (which is typically less than the bare electric charge of the particles 
\cite{Kennedy2003, Morfill2009}), and the dielectric permittivity $\epsilon$ of the embedding medium. In case of dusty plasmas, 
$\epsilon$ is equal to the dielectric permittivity of vacuum, $\epsilon_0$, for all purposes of the present article in which we 
adhere to SI units. The screening parameter, $\kappa_{Y} = 1 / (\lambda_{Y} \sqrt{n})$, is the normalized inverse of the Debye 
screening length $\lambda_{Y}$, which depends on the ion population. In an embedding plasma that consists of neutral particles and 
univalent positive ions only, $\lambda_{Y} = \sqrt{\epsilon k_B T / (e^2 n_i)}$ is the Debye length in terms of the proton 
elementary charge $e$, and the unperturbed (3D) ion number density $n_i$ of the ions far from the charged particle's confining 
plane.

The Yukawa model in two dimensions is best realized experimentally for charged colloids which are confined between two highly 
charged glass plates \cite{Chang1988,Lowen92,Palberg05}. There, the screening is caused by the microions between the plates 
\cite{Chang1988} and it can be tuned by adding salt. The experimentally observed freezing phase sequence has been found to agree 
with the theoretical predictions assuming a 2D Yukawa interaction \cite{Palberg05}.

\subsection{Kompaneets monolayer}

\begin{figure}[hbt]
\includegraphics[width=0.95\columnwidth]{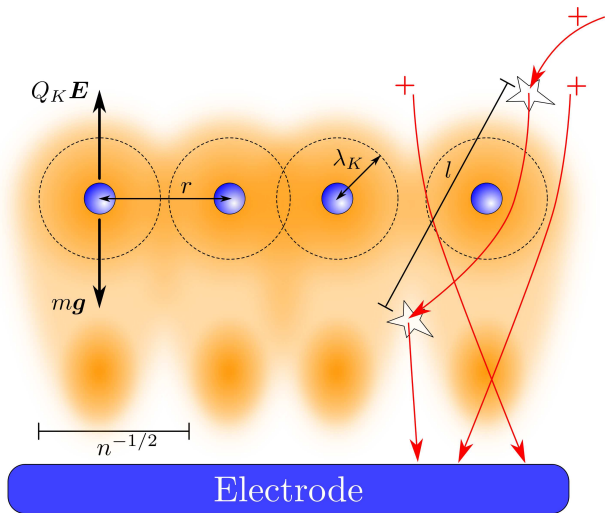}
\caption{\label{fig:Kompaneets_schematic} Edge-on schematic of a dusty plasma monolayer. Negatively charged dust particles (filled 
circles) levitate in a well-defined 2D layer above an electrode in a radio frequency discharge plasma chamber, at a height where 
gravity is balanced by the vertical electrostatic force. The mean distribution of ions is color-coded. Three characteristic ion 
trajectories are sketched by arrows signed with $+$. Subsequent collisions between ions and neutral particles are separated on 
average by the ion-neutral mean free path $l$. Ions are focused in the downstream direction below the dust particles, giving rise 
to positive space-charges in the plasma wake region. The effective dust particle interactions are quantified by the three 
dimensionless parameters $\Gamma_K$, $\kappa_{K} = 1 / (\lambda_{K} \sqrt{n})$, and $\zeta = \lambda_K / l$.}
\end{figure}

The second class of systems studied in this article is schematically depicted in Fig.~\ref{fig:Kompaneets_schematic}. A radio 
frequency discharge chamber contains a weakly ionized plasma (of neutral gas particles, electrons, and ions), and negatively 
charged dust particles are levitating atop an electrode on the bottom of the chamber. Confinement of the dust particles to a 
well-defined 2D layer is achieved by a force balance between gravitation and electrostatic repulsion. Unlike the particles in the 
spatially unbounded Yukawa system, the ions in the radio frequency chamber exhibit a highly non-equilibrium steady state with a 
non-zero plasma current towards the electrode, where positive ions are adsorbed. Attraction between dust particles and ions causes 
downstream focusing of ions in the so-called plasma wake region. As a consequence, every dust particle trails a positive 
space-charge in the downstream direction, which causes the effective pair-potential between charged dust particles to be 
anisotropic in 3D.

Kompaneets \textit{et al.} \cite{Kompaneets2007} have presented a self-consistent steady-state solution for the effective pair 
potential between the dust particles, taking into account the external electric field $\textbf{E}$ towards the electrode, and the 
collisions between ions and electrically neutral particles in the plasma. The resulting effective particle pair-potential, 
obtained under the assumptions of the mobility-limited ion drift in the field $\textbf{E}$, velocity-independent ion-neutral 
scattering cross section, and further assumptions that are outlined in the original reference, has been derived and described 
comprehensively in Ref.~\cite{Kompaneets2007}. We will refer to this kinetic pair-potential as the Kompaneets pair potential. For 
particles that are perfectly confined to a plane perpendicular to the plasma current, the in-plane Kompaneets potential $U_K(x)$ 
is given by
\begin{equation}\label{eq:K_pot}
\begin{split}
\frac{U_K(x)}{k_\text{B}T} =&~\Gamma_K~\dfrac{2\zeta\kappa_{K}}{\pi}~\text{Re}\int_0^\infty \frac{dt}{1+\zeta^{-2} Y(t)}\\
&\times K_0\left(x\zeta\kappa_{K}\sqrt{\frac{t^2+\zeta^{-2} X(t)}{1+\zeta^{-2} Y(t)}} \right),
\end{split}
\end{equation}
where $K_0$ is the zeroth-order modified Bessel function of the second kind \cite{Abramowitz}, and the two auxiliary 
functions $X(t)$ and $Y(t)$ are defined as
\begin{equation}
\begin{split}
&X(t)=1-\sqrt{1+it},\\
&Y(t)=\frac{2\sqrt{1+it}}{it}\int_0^1\frac{d\alpha}{[1+it(1-\alpha^2)]^2}-\frac{1}{it(1+it)}.
\end{split}
\end{equation}
In Eq.~\eqref{eq:K_pot}, the prefactor $\Gamma_K = Q_K^2 \sqrt{n} / (4 \pi \epsilon k_\text{B} T)$ quantifies the interaction 
strength in terms of the effective charge $Q_{K}$, and the screening parameter is defined as $\kappa_{K} = 1 / (\lambda_{K} 
\sqrt{n})$, where $\lambda_K = \sqrt{\epsilon El / (e^2 n_i)}$ is a field-induced screening length. In addition, the Kompaneets 
potential depends on the collision parameter $\zeta=\lambda_{K}/l$, where $l$ is the mean free path between two consecutive 
collisions of an ion and neutral gas particles (``ion-neutral mean free path'', for short).

For close-contact configurations ($r \ll \zeta^s\lambda_K$, where $1/3\leq s\leq1$, depending on the magnitude of $\zeta$ 
\cite{Kompaneets2007}), the Kompaneets potential tends to the bare Coulomb potential:
\begin{equation}\label{eq:zeta_0_K}
x\ll\zeta^s/\kappa_{K}:\qquad \frac{U_K(x)}{k_\text{B}T} \to \dfrac{\Gamma_K}{x}.
\end{equation}
Hence, the Coulomb potential is recovered at all distances $x$ in the limit $\zeta \to \infty$, corresponding to a very large 
field $E$, or a very small ion mean free path $l$ or/and ion density $n_i$. For large particle separations and finite values of 
$\zeta$, the Kompaneets potential reduces to its in-plane asymptotic form
\begin{equation}\label{eq:Komlimit}
\left.\frac{U_K(x)}{k_\text{B}T}\right|_{x \to \infty} = \dfrac{\Gamma_K}{6\sqrt{2}\kappa_{K}^2x^3} \left(60 \zeta^2 - 1\right)
+\mathcal{O}(x^{-4}).
\end{equation}
The leading order asymptotic form of the anisotropic out-of-plane electrostatic potential is proportional to $x^{-2}$, and is 
given in Eq.~(8) of Ref.~\cite{Kompaneets2007} (in Gaussian units).

In typical dusty plasma experiments the effective interaction potential can be measured for particle distances $x \approx 1$ 
(\textit{i.e.}, close to the mean geometric distance) by particle video tracking \cite{Konopka2000}. It has been shown in 
Ref.~\cite{Kompaneets2007}, that the pair-potential in the experimentally directly accessible narrow range of particle separations 
can be fitted equally well by the Yukawa as well as the Kompaneets form. However, one should expect that the qualitative 
differences between the Yukawa and Kompaneets potentials, most particularly in their long-ranged asymptotic forms, can have a 
considerable influence on collective dynamics \cite{Roeker2012} and phase transitions.

\section{Mode Coupling Theory}\label{sec:MCT}

The glassy state is characterized by liquid-like static pair correlations without long range order, and a non-zero value of the 
non-ergodicity parameter $f_q = \lim_{t\rightarrow \infty} \phi_q(t)$, which is the long time limit of the wavenumber- and 
time-dependent autocorrelation function $\phi_q(t)$ of the number density. The parameter $f_q$ is also called the form factor or 
the Debye-Waller factor. In contrast to the glassy state, the liquid state is characterized by a vanishing non-ergodicity 
parameter, $f_q = 0$, for all wavenumbers $q$. In MCT, $f_q$ is calculated as \cite{Bengtzelius1984, Goetze2009}
\begin{equation}\label{eq:trans}
 \frac{f_q}{1-f_q}=\mathcal{F}_q [f],
\end{equation}
where in 2D \cite{Bayer2007}
\begin{equation}\label{eq:kernel}
 \mathcal{F}_q [f]= \frac{S_q}{8 \pi^2 q^4} \int d^2k~S_k S_p
\left(\mathbf{q}\cdot\mathbf{k}~c_k+\mathbf{q}\cdot\mathbf{p}~c_p\right)^2 f_k f_p,
\end{equation}
with $\mathbf{p}=\mathbf{q}-\mathbf{k}$.

The static structure factor $S_{q}$ and direct correlation function $c_{q} = 1 - 1 / S_{q}$ are the only input to the MCT 
equations, conveying information about the particle interactions. Note that the number density $n$ does not explicitly enter into 
Eq.~\eqref{eq:kernel}, since all lengths and wave vectors are expressed in units of $1/\sqrt{n}$ and $\sqrt{n}$, respectively: In 
our notation the wave vector $\boldsymbol{q}$ is the dimensionless Fourier conjugate variable to the dimensionless distance vector 
$\boldsymbol{x} = \boldsymbol{r} \sqrt{n}$.

The Lamb-M\"ossbauer factor $f^s_q=\lim_{t\rightarrow \infty} \phi^s_q(t)$, which is the long-time limit of the wavenumber- and 
time-dependent, Fourier transformed tagged particle position autocorrelation function $\phi^s_q(t)$, is calculated in MCT 
according to \cite{Fuchs1998}
\begin{equation}\label{eq:transs}
 \frac{f^s_q}{1-f^s_q}=\mathcal{F}^s_q [f,f^s],
\end{equation}
where \cite{Bayer2007}
\begin{equation}\label{eq:kernels}
 \mathcal{F}^s_q [f,f^s]= \frac{1}{4\pi^2 q^4} \int d^2k~S_k (\mathbf{q}\cdot \mathbf{k})^2 c_k^2 f_k f^s_p
\end{equation}
and, once again, $\mathbf{p}=\mathbf{q}-\mathbf{k}$.

\begin{figure}[htb]
\includegraphics[width=0.74\columnwidth, angle=-90]{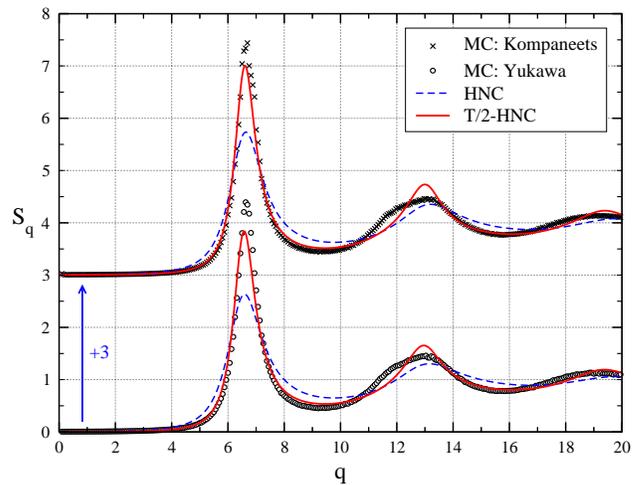}
\caption{\label{fig:Sq_MC_HNC_ThalfHNC}Static structure factors for a Yukawa monolayer with $\Gamma_Y = 100$ and $\kappa_Y = 2.0$ 
(lower three data sets) and a Kompaneets monolayer with $\Gamma_K = 300$, $\kappa_K = 2.0$, and $\zeta = 0.25$ (upper three 
datasets). Crosses and circles: Monte Carlo simulation results. Dashed curves: HNC integral equation solution. Solid curves: 
Solution of the T/2-HNC in Eq.~\eqref{eq:ThalfHNC}. The Kompaneets monolayer structure factors are shifted by $3$ units along the 
vertical axis for clarity. }
\end{figure}

We evaluate the integrals in Eqs.~\eqref{eq:kernel} and \eqref{eq:kernels} numerically and solve Eqs.~\eqref{eq:trans} and 
\eqref{eq:transs} iteratively with iteration seeds ${f_q}^0 = {f^s_q}^0 = 1$ \cite{Franosch1997}. To evaluate the integrals 
numerically we use $N=200$ equidistant grid points with spacing $\Delta q= 0.2$, minimal wavenumber $q_\text{min}= 0.1$ and 
maximal wavenumber $q_\text{max}= 39.9$. Test calculations with $N=500$ grid points allow us to estimate the numerical error due 
to integral discretization, which is around $5\%$ in the glass transition temperatures. The static structure factor is obtained 
from the (Fourier transformed) solution of the T/2-HNC integral equation \cite{Hajnal2011}
\begin{equation}\label{eq:ThalfHNC}
\gamma(x) = \int d^2x'~c(|\mathbf{x}-\mathbf{x}'|)
\left[ \exp\left\lbrace \gamma(x') - \dfrac{\displaystyle{2 U(x')}}{\displaystyle{k_B T}}\right\rbrace - 1 \right],
\end{equation}
for an isotropic 2D fluid in terms of the indirect and direct correlation functions $\gamma(x)$ and $c(x)$ 
\cite{Hansen}. Equation~\eqref{eq:ThalfHNC} is solved by means of a numerical spectral solver for liquid integral 
equations in an arbitrary number of spatial dimensions, which has been comprehensively described in Ref.~\cite{Heinen2014}, and 
which is based on methods that have been originally introduced in Refs.~\cite{Ng1974, Talman1978, Rossky1980, Hamilton2000, 
Hamilton}. The spectral solver operates on logarithmically spaced grids of wavenumbers and radii, providing a 
high-resolution structure factor that is mapped to the above mentioned equidistant wavenumber grid by quadratic interpolation.

Note that Eq.~\eqref{eq:ThalfHNC} is a simple modification of the well-known hypernetted chain (HNC) integral equation 
\cite{Morita1958,Hansen}, which is recovered when the term $2 U(x')$ in the integrand is replaced by $U(x')$. Thus, 
the solution of the T/2-HNC equation coincides exactly with the solution of the HNC integral equation for a system in which the 
temperature has been scaled down by a factor of $1/2$. In Ref.~\cite{Hajnal2011}, the MCT glass transition was studied for 
two-dimensional binary mixtures of aligned point-dipoles with a long-ranged repulsive pair potential that is proportional to the 
inverse cube of the particle separation, $r^{-3}$. It was empirically found in Ref.~\cite{Hajnal2011} that the T/2-HNC scheme 
predicts the static structure factors of the strongly repulsive 2D binary dipole mixtures with a significantly higher accuracy 
than the HNC scheme. In order to test the accuracy of the T/2-HNC scheme for the Yukawa- and Kompaneets monolayer systems, we have 
simulated 2D equilibrium liquids with strong repulsive pair potentials of both types, and compared the static structure factor 
from the simulation to the HNC and the T/2-HNC scheme solutions. Our results, shown in Fig.~\ref{fig:Sq_MC_HNC_ThalfHNC}, underpin 
the good accuracy of the T/2-HNC and its supremacy over the HNC scheme. We have obtained the datasets represented by crosses and 
circles in Fig.~\ref{fig:Sq_MC_HNC_ThalfHNC} from Metropolis Monte Carlo (MC) simulations in the $NLT$-ensemble of constant 
particle number $N$, constant system area $L^2$, and constant temperature $T$. A square simulation box with periodic boundary 
conditions in both Cartesian directions was used in our simulations, and we haven chosen the parameters $\Gamma_Y = 100$ and 
$\kappa_Y = 2.0$ for the Yukawa monolayer of $N = 10.000$ particles and $\Gamma_K = 300$, $\kappa_K = 2.0$, and $\zeta = 0.25$ for 
the Kompaneets monolayer of $N = 12.000$ particles. Both simulated systems are strongly coupled equilibrium liquids not far from 
the crystal-liquid transition point. In our MC simulations, the direct particle interactions are truncated at a dimensionless 
cutoff radius of $x_c = r_c \sqrt{n} = 5$ in case of Yukawa interactions, and at $x_c = 12.5$ in case of Kompaneets interactions. 
For pair separations $x > x_c$, the pair-potential is set equal to zero in the simulations. Varying its numerical value, we have 
checked that the cutoff radius is large enough and does not have a significant effect on the measured quantity $S_q$.

Note from Fig.~\ref{fig:Sq_MC_HNC_ThalfHNC} that despite its improved accuracy in comparison to the standard HNC scheme, the 
T/2-HNC scheme still tends to underestimate the principal peak height in $S_q$. In addition note that we apply the T/2-HNC scheme 
in the following sections to systems at the liquid-glass transition, that is, beyond the equilibrium fluid regime for which the 
accuracy of the integral equation scheme can be tested by comparison to crystallization-free simulations. Moreover, the 
approximate T/2-HNC scheme is combined in the following with the approximate MCT equations. The combined uncertainty of the 
resulting glass transition lines cannot be easily estimated and, thus, the numerical values of the glass transition temperatures 
must be taken with some caution. Nevertheless, the dominating qualitative features of $S_q$ are contained in the T/2-HNC solution, 
and the features of the glass transition curves can be expected to be at least qualitatively correct.

It is important to note also that the T/2-HNC scheme is empirically justified only in case of strong enough particle interactions. 
In the limit of vanishing interactions, $\Gamma_Y \to 0$ or $\Gamma_K \to 0$, the T/2-HNC scheme predicts twice the correct 
asymptote $c(r) \to \exp\left\lbrace-U(r)/k_B T\right\rbrace - 1$ for the direct correlation function (i.e., twice the Mayer 
function). A related issue is the wrong long-distance decay -- the T/2-HNC scheme yields twice the correct expression 
$\lim_{r\to\infty} c(r) = -U(r) / k_B T$. This wrong long-distance decay is observed for all values of the potential prefactor.

\section{Results}\label{sec:Results}

\subsection{Glass transition diagrams}

\begin{figure}
\includegraphics[width=0.72\columnwidth, angle=-90]{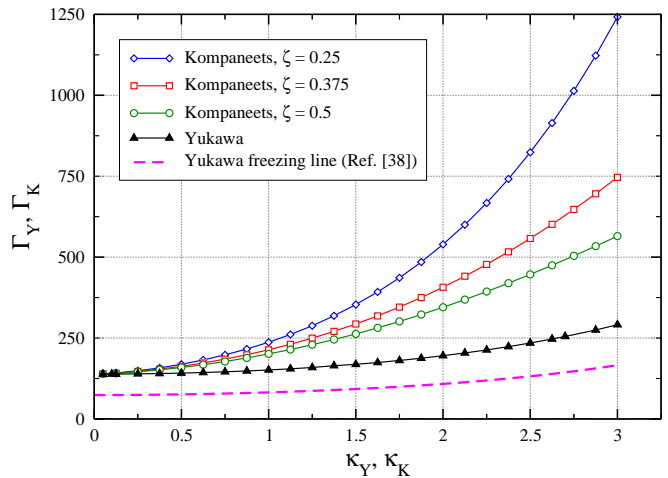}
\caption{\label{fig:Kompaneets}Glass transition curves in the \mbox{($\kappa, \Gamma$)}-plane, for the Yukawa potential (black 
curve with triangles) and three different Kompaneets potentials with parameters $\zeta = 0.25$ (solid curves with diamonds), 
$\zeta = 0.375$ (solid curve with squares) and $\zeta = 0.5$ (solid curve with circles). The dashed curve is the 2D Yukawa 
freezing line from Ref.~\cite{Hartmann2005}. }
\end{figure}
The glass transition curves for the Yukawa monolayer, and for three different Kompaneets-monolayers with different values of the 
collision parameter $\zeta$, are shown in the transition diagram in Fig.~\ref{fig:Kompaneets}. In the transition diagram, the 
screening parameters $\kappa_{Y}$ and $\kappa_{K}$ vary along the horizontal axis, and the coupling parameters $\Gamma_Y$ and 
$\Gamma_K$ vary along the vertical axis. The data points in Fig.~\ref{fig:Kompaneets} represent the lowermost values of 
$\Gamma_{Y,K}$ for which $f_q$ assumes a non-zero value at given values of $\kappa_{Y,K}$. Note that at the glass transition, 
$f_q>0$ for all finite values of $q$ smaller than the $q$ cutoff. In our implementation, we have tested $f_q$ at $q = 3.9$ to 
identify the glass transition points. In the One-Component-Plasma (OCP) limit $\kappa_{Y,K} \to 0$, both the Yukawa and the 
Kompaneets potential reduce to the unscreened Coulomb potential [see Eqs.~\eqref{eq:Y_pot} and \eqref{eq:zeta_0_K}], and the glass 
transition curves close in on the T/2-HNC-MCT approximation for the OCP glass transition point, $\Gamma_Y = \Gamma_K = 138.5$.

While the glass transition curves are qualitatively similar for the Yukawa and the Kompaneets systems, the transition occurs at 
higher values of the coupling parameter in case of the Kompaneets monolayer. For decreasing values of the parameter $\zeta$, the 
differences between the Yukawa and Kompaneets glass transition curves are increasing. Such trend is not surprising: As we discuss 
in the next subsection (see also Fig.~\ref{fig:potential}), the deviation of the Kompaneets potential from the Yukawa-like form 
drastically increases as $\zeta$ decreases. On the other hand, in the limit $\zeta \to \infty$ the Kompaneets potential tends to 
the Coulomb form, so the Kompaneets glass transition curve in this case would be a horizontal line in the transition diagram of 
height $\Gamma_K = 138.5$.

For 3D Yukawa systems it has been found that the glass transition and crystallization (freezing) lines are approximately parallel 
in the $(\kappa_Y, \Gamma_Y)$-plane \cite{Yazdi2014}. The same similarity between the glass transition and crystallization lines 
is found for the 2D Yukawa monolayer in Fig.~\ref{fig:Kompaneets}, where we plot the 2D Yukawa freezing line reproduced from 
Ref.~\cite{Hartmann2005} (dashed curve), and the T/2-HNC-MCT 2D Yukawa glass transition line (black curve with triangles). In 
Ref.~\cite{Hartmann2005}, the crystallization line was obtained from simulations, and it was approximated by the inverse 
polynomial $\Gamma = \Gamma^{\ast} / (1 + f_2 \kappa^2 + f_3 \kappa^3 + f_4 \kappa^4$), with $\Gamma^{\ast}= 73.9 = 
131/\sqrt{\pi}$, $f_2=-0.1235$, $f_3=0.0248$ and $f_4=-0.0014$. Note that the 2D ion-sphere radius $a = 1/\sqrt{\pi n}$ (also 
called Wigner-Seitz radius) was used as a unit of length in Ref~\cite{Hartmann2005}, instead of the mean geometric distance 
$1/\sqrt{n}$ utilized in the present paper. Therefore, one has to take account of a $1/\sqrt{\pi}$ prefactor difference in the 
definitions of the Yukawa coupling parameter and the inverse Yukawa screening parameter.

\subsection{Potentials and structure factors at the glass transition}

\begin{figure}
\includegraphics[width=0.72\columnwidth, angle=-90]{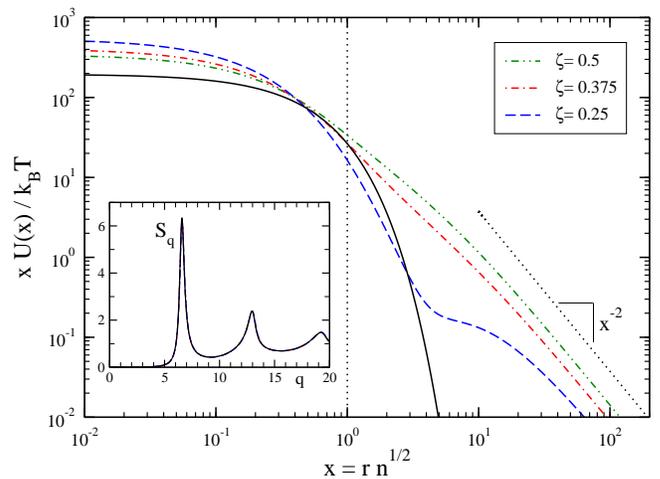}
\caption{\label{fig:potential}Effective pair-potentials for $\kappa_Y = \kappa_K = 2.0$, for values of $\Gamma_Y$ and $\Gamma_K$ 
at the liquid-glass transition point. Solid curve: Yukawa potential for $\Gamma_Y = 195.4$. Dashed, dot-dashed and dot-dot-dashed 
curves: Kompaneets potentials for $\zeta = 0.25, 0.375$ and $0.5$, and $\Gamma_K = 539.4, 406.8$ and $345.0$, respectively. All 
potentials are multiplied by their argument, $x=r\sqrt{n}$, to expose the differences. The inset features the corresponding static 
structure factors $S_q$ in T/2-HNC approximation. All four functions $S_q$ are overlapping on the scale of the inset. The 
principal peak heights of the structure factors are $S_q = 6.33$ for the Yukawa system, and $S_q = 6.26, 6.23$ and $6.19$ for the 
Kompaneets systems with $\zeta = 0.5, 0.375$ and $0.25$, respectively. }
\end{figure}

In Fig.~\ref{fig:potential} we plot the Yukawa potential and three different in-plane Kompaneets potentials for different values 
of $\zeta$, all at the glass transition for $\kappa_Y = \kappa_K = 2.0$. The full set of parameters, including the glass 
transition values of $\Gamma_Y$ and $\Gamma_K$, is provided in the figure caption. Note that the potentials in 
Fig.~\ref{fig:potential} are multiplied by their argument $x = r\sqrt{n}$, to expose the differences. The curves corresponding to 
Kompaneets pair-potentials (with $r^{-3}$ asymptotics) therefore decay proportionally to $x^{-2}$ for large values of $x$. The 
inset of Fig.~\ref{fig:potential} features the T/2-HNC static structure factors $S_q$, corresponding to the four different 
potentials plotted in the figure's main panel. Despite the pronounced differences between the four potentials (in particular 
around the most frequently sampled mean geometric distance $x = 1$), all four functions $S_q$ at the glass transition are 
indistinguishable on the scale of the figure inset. The principal peak heights of the four structure factors differ only slightly 
in their values.

\subsection{A fallacious re-entrant state sequence}

As pointed out in Sec.~\ref{sec:Model_Systems}, the 2D Yukawa model with its many simplifying assumptions is merely a toy model 
for experimentally observable monolayers of mesoscopic charged particles. For the important class of ground-based dusty plasma 
experiments the kinetic pair-potentials are far more realistic, and among them the Kompaneets pair potential stands out with its 
realistic model assumptions. In this section, we allude to the possible consequences of an over-simplified interpretation of 
charged particle monolayers in terms of the Yukawa model. We show that the liquid-glass transition of a system with 
Kompaneets-like pair potential appears as a non-monotonic curve (corresponding to liquid-glass-liquid state re-entrance) when it 
is plotted in terms of the inappropriate parameters of the Yukawa potentials that represent a best fit to the actual (Kompaneets) 
potential around the mean geometric distance $x = r\sqrt{n} = 1$.

\begin{figure}[htb]
\includegraphics[width=0.72\columnwidth, angle=-90]{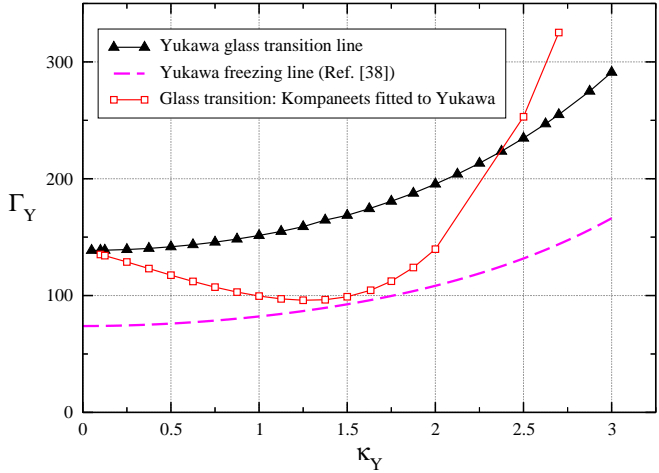}
\caption{\label{fig:Yukawa} Kompaneets and Yukawa glass transitions in the Yukawa parameter plane. The curve with triangles: 
Yukawa glass transition curve in the Yukawa screening- and coupling-parameter \mbox{($\kappa_Y, \Gamma_Y$)}-plane. The curve with 
squares is the glass transition curve for Kompaneets pair-potentials which have been optimally fitted to the corresponding Yukawa 
potential in the region $0.7 < x < 3$, by pointwise tuning of the Kompaneets screening parameter $\kappa_K$ and coupling parameter 
$\Gamma_K$. The parameters $l = 2.3$ mm and $n = 1.5$ mm$^{-2}$ are held fixed for the Kompaneets potential. The dashed curve is 
the 2D Yukawa freezing line from Ref.~\cite{Hartmann2005}.}
\end{figure}

In Fig.~\ref{fig:Yukawa} we plot the Yukawa glass transition curve that is also shown in Fig.~\ref{fig:Kompaneets} (black curves 
with triangles). The 2D Yukawa freezing line, reproduced from Ref.~\cite{Hartmann2005}, is also shown (dashed line) to allow a 
better comparison to the glass transition line than on the scale of Fig.~\ref{fig:Kompaneets}. The curve with open squares in 
Fig.~\ref{fig:Yukawa} is generated as follows: For given values of the two Yukawa parameters $\kappa_Y$ and $\Gamma_Y$, we 
calculate the Kompaneets potential that fits best to the Yukawa potential in the distance range $0.7 < x < 3$ which is most 
frequently sampled by the particles \cite{Konopka2000}. The fit is conducted as follows: For given values of $l$ and $n$, which 
yields the combination $\zeta\kappa_{K}\equiv(l \sqrt{n})^{-1}$ ($\simeq0.354$ for the example shown in the figure), we tune the 
two remaining, independent Kompaneets parameters $\kappa_K$ and $\Gamma_K$; an optimal fit is achieved by minimizing the square 
deviation $\int_{0.7}^3 dx {[U_Y(x) - U_K(x)]}^2$ between the two potentials. We then calculate $S_q$ for the best-fitting 
Kompaneets potential in the T/2-HNC scheme, and use it as the input to the MCT equations \eqref{eq:trans} and \eqref{eq:kernel} 
for $f_q$. If $f_q = 0$, the system is classified as liquid, and if $f_q > 0$, it is classified to be in the glassy state. We 
repeat the full procedure for various Yukawa parameters $\kappa_Y$ and $\Gamma_Y$, which are tuned by interval bisection, until we 
find for each $\kappa_Y$ the smallest (critical) value of $\Gamma_Y$ at which the best-fitting Kompaneets system vitrifies.

Thus, the curve with open squares in Fig.~\ref{fig:Yukawa} is the glass transition curve of a dusty plasma monolayer with 
Kompaneets-like interactions, as it would appear when plotted in terms of the dimensionless parameters $\kappa_Y$ and $\Gamma_Y$ 
of the Yukawa potentials that best fit the actual Kompaneets potential around the mean geometric distance, where the potential is 
directly accessible \cite{Konopka2000}. Therefore, if one observes vitrification in a dusty plasma monolayer and assumes 
Yukawa-like interactions in the experiment analysis, the transition behavior may be misinterpreted as a re-entrant 
liquid-glass-liquid state sequence, while the transition diagram in terms of the three relevant Kompaneets potential parameters 
does not exhibit any re-entrance (see Fig.~\ref{fig:Kompaneets}).

\subsection{Non-ergodicity parameters}

\begin{figure}[htb]
\includegraphics[width=0.72\columnwidth, angle=-90]{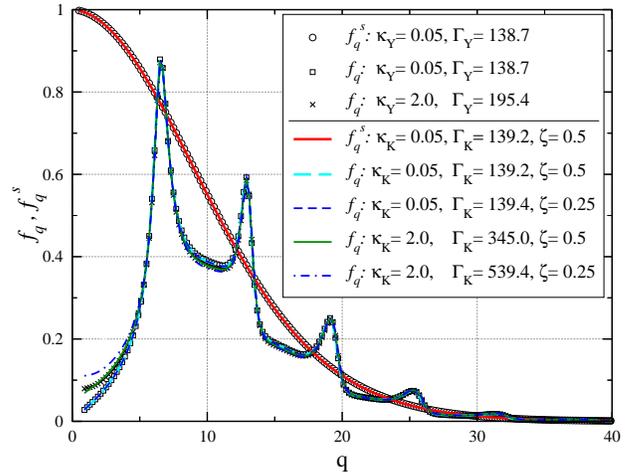}
\caption{\label{fig:fq} The Lamb-M\"ossbauer factors $f^s_q$ (monotonically decaying as functions of $q$) and the form factors 
$f_q$ (non-monotonic functions of $q$) for various Yukawa monolayers (symbols) and Kompaneets monolayers (curves) at their 
respective MCT glass transition points, with potential parameters as indicated in the legend.}
\end{figure}
We turn our attention now to the $q$-dependent Debye-Waller and Lamb-M\"ossbauer factors at the glass transition, which are 
plotted in Fig.~\ref{fig:fq}. It is observed that $f_q$ approaches zero in the limit $q \to \infty$. In the opposite limit $q \to 
0$, the functions $f_q$ for Yukawa- and Kompaneets potentials with a small value of the screening parameter $\kappa_Y = \kappa_K = 
0.05$ assume very small but non-zero values, and for finite wavenumbers $q$, all plotted functions $f_q$ deviate clearly from 
zero. Finite wavelength density modulations, cannot relax in the glassy state since this would require a collective rearrangement 
of particles on the length scale of some nearest neighbor cage diameters. Observing Fig.~\ref{fig:fq} and the inset of 
Fig.~\ref{fig:potential}, one can see that the most resilient density modulations (corresponding to the principal maximum in 
$f_q$) are for $q \approx 2\pi$, that is, at the wavenumber that corresponds to the static structure factor principal maximum, and 
to the Fourier conjugate of the nearest neighbor (mean geometric) distance. Very long wavelength ($q\to0$) density modulations 
cannot relax in the glassy state in general, as indicated by the finite values of $f_{q\to0}$ in Fig.~\ref{fig:fq}. This is due to 
the finite isothermal compressibility, $S_{q\to0}$, of the system. Note here in particular that the Debye-Waller factor of 
hard-sphere system remains finite as $q$ approaches zero \cite{Franosch1997}, which is in line with the rather large 
compressibility of such hard-sphere system. However, in the OCP limits $\kappa_Y \to 0$ and $\kappa_K \to 0$, in which both the 
Yukawa and the Kompaneets potential reduce to the Coulomb potential, the isothermal osmotic compressibility coefficient vanishes 
\cite{Baus1980, Hansen}, which corresponds to an infinite thermodynamic driving force for the leveling of 
long-wavelength density modulations. Therefore, $f_{q\to0}=0$ in the OCP limit.

The Yukawa- and Kompaneets system are indistinguishable in the OCP limit. This facilitates computation of the OCP-limiting 
behavior in terms of a small-q asymptotic expansion of the Yukawa monolayer Debye-Waller factor. As demonstrated in the Appendix, 
the T/2-HNC solution for the direct correlation function of a Yukawa monolayer can be approximated as
\begin{gather}\label{eq:smallq_cq_ThalfHNC_2dYuk}
c_q \approx -\dfrac{2 U_Y(q)}{k_B T} = - \dfrac{4\pi \Gamma_{Y}}{\sqrt{\kappa_{Y}^2 + q^2}}~~~\text{for}~q + q_t \gg\kappa_Y^2+q^2,
\end{gather}
that is, when both the wavenumber $q$ and the screening parameter $\kappa_Y$ are small and within a certain ratio of each other. 
In Eq.~\eqref{eq:smallq_cq_ThalfHNC_2dYuk}, $q_t$ is a dimensionless non-negative threshold wavenumber with a typical value of 
$q_t \sim 0.1$. The corresponding small-$q$, small-$\kappa_Y$ form of $\mathcal{F}_{q}$ is obtained from a functional Taylor 
expansion \cite{Bayer2007,Yazdi2014}, resulting in
\begin{equation}\label{eq:kernel_sq}
\mathcal{F}_{q} = (\alpha + \beta q^2 \ldots)S_{q}
\end{equation}
where $S_q=(1-c_q)^{-1}$,
\begin{equation}\label{eq:alpha}
\alpha = \frac{1}{4\pi} \int~dk~k S_k^2 \left(c_k^2 + k c_k  c_k'+ \frac{3}{8} k^2 {c'}_k^2\right) f_k^2
\end{equation}
and
\begin{equation}
\begin{split}\label{eq:beta}
\beta= &\frac{1}{8\pi} \int~dk~k S_k^2 \left({c'_k}^2+\frac{5}{32} k^2 {c''_k}^2+\frac{3}{2} k c'_k c''_k\right.\\
&\left.+c_k c''_k+\frac{1}{4}k c_k c'''_k+\frac{5}{24} k^2 c'_k c'''_k \right) f_k^2.
\end{split}
\end{equation}
Considering only the leading order of the approximation, this translates into the small-$q$, small-$\kappa_Y$ limiting behavior of 
the Yukawa monolayer Debye-Waller factor,
\begin{equation}\label{eq:DW_lowq_finitescreening}
f_q = {\alpha\left[ 1 + \alpha+ \dfrac{4\pi \Gamma_{Y}}{\sqrt{\kappa_{Y}^2 + q^2}} \right]}^{-1}~~~\text{for}~q + q_t \gg 
\kappa_Y^2 + q^2,
\end{equation}
in T/2-HNC approximation. For finite $\kappa_Y$, the function $f_q$ in Eq.~\eqref{eq:DW_lowq_finitescreening} assumes a positive 
value for $q=0$, and increases $\propto q^2$ when $q\to0$. Only in the OCP limit $\kappa_Y = 0$, the function $f_q$ in 
Eq.~\eqref{eq:DW_lowq_finitescreening} vanishes for $q=0$, and increases initially as $\propto q$. In a broad scale the $f_q$ 
asymptotic for $\kappa_Y= 0.05$ and $\Gamma_Y= 138.7$ is almost linear.

Note here that the small-$q$ limiting OCP Debye-Waller factor is qualitatively different in two and three dimensions. In 3D, the 
function $f_q$ vanishes in the OCP limit $\kappa_Y=0$ as $\propto q^2$ \cite{Yazdi2014}. In contrast to $f_q$, the 
Lamb-M\"ossbauer factor $f^s_q$ in MCT approximation does not critically depend on the form of the pair potential, since 
$\mathcal{F}^s_q$ in Eq.~(\ref{eq:kernels}) and also the small-$q$ limit of $\mathcal{F}^s_q$ \cite{Bayer2007} do not depend on 
$S_q$.

\section{Conclusions}\label{sec:Conclusions}

We have calculated the liquid-glass transition boundaries in the state diagram spanned by the screening parameters and coupling 
parameters of 2D monolayers with Yukawa- and Kompaneets-like pair potentials, in T/2-HNC-MCT approximation. While both types of 
systems exhibit qualitatively similar glass transition curves, there is a quantitative difference in the vitrification 
temperature, which decreases as a function of the collision parameter $\zeta$ of the Kompaneets pair-potential. Both the 
Kompaneets- and Yukawa-monolayer reduce to a two-dimensionally confined OCP in the limit of infinite $\lambda_Y$ and $\lambda_K$.

In contrast to the over-simplifying 2D Yukawa model, the kinetic pair-potentials, including in particular the Kompaneets 
pair-potential, provide far more accurate descriptions of the interactions between dust grains in typical ground-based complex 
plasma experiments. We have demonstrated that a glass transition in a dusty plasma monolayer is prone to a qualitative 
misinterpretation if the simple Yukawa model is invoked in its analysis: While the glass transition line is a monotonic function 
in terms of the three relevant, dimensionless Kompaneets pair-potential parameters, it appears to be non-monotonic corresponding 
to a fallacious liquid-glass-liquid re-entrance when the pair interactions are misinterpreted as Yukawa-type interactions.

A promising task for future research would be the generalization of our results to binary systems, in order to understand the 
different glass types in mixtures. Since the crystalline states in such systems are pretty complex \cite{Messina08}, the glass 
transition scenarios are also expected to be much more complex than in the monodisperse system.

\section*{Acknowledgements}

The authors acknowledge support from the European Research Council under the European Union's Seventh Framework Programme, Grant 
Agreement No.~267499. M.H. acknowledges support by a fellowship within the Postdoc-Program of the German Academic Exchange Service 
(DAAD).

\section*{Appendix}

\begin{figure}[htb]
\includegraphics[width=0.72\columnwidth, angle=-90]{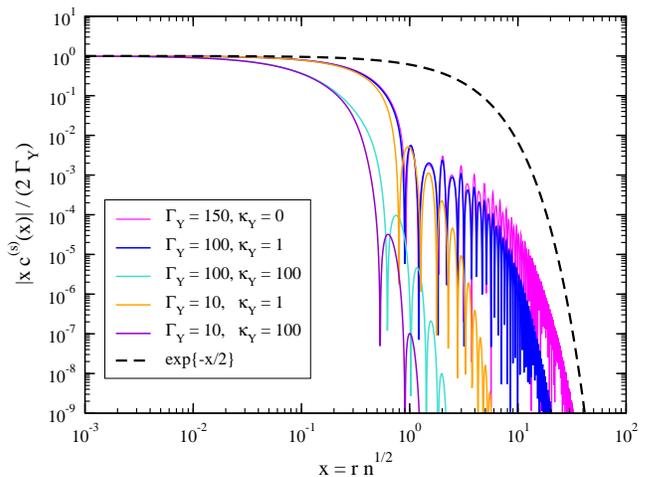}
\caption{\label{fig:cx_decay}Function $|x c^{(s)}(x)|$ for Yukawa monolayers and various potential parameters, calculated in 
T/2-HNC approximation, is bound from above by $2\Gamma_Y \exp\left\lbrace-x/2\right\rbrace$.}
\end{figure}
Here we validate the small-$q$, small-$\kappa_Y$ result for the T/2-HNC direct correlation function of a Yukawa monolayer in 
Eq.~\eqref{eq:smallq_cq_ThalfHNC_2dYuk}. We begin by noting that the function $c(x)$ can be split into the sum
\begin{equation}\label{eq:c_short}
c(x) = c^{(s)}(x) - 2 \Gamma_Y \dfrac{\exp(-\kappa_Y x)}{x},
\end{equation}
of a short-ranged part, $c^{(s)}(x)$, and the asymptotic long-ranged part, $-2\Gamma_Y \exp(-\kappa_Y x)/x$. In 
Eq.~\eqref{eq:c_short}, the peculiar long-ranged asymptotics $c(x\to\infty) = -2U(x)/{k_B T}$ of the T/2-HNC scheme solution has 
been taken into account (\textit{c.f.}, our discussion at the end of Sec.~\ref{sec:MCT}). The direct correlation function in 
wavenumber-space is calculated as the isotropic 2D Fourier transform (Hankel transform)
\begin{eqnarray}\label{eq:c_qnull_int}
c_{q} &&= 2 \pi \int\limits_0^\infty~dx~xc(x) J_0(qx)\\
&& = \underbrace{2 \pi \int\limits_0^\infty~dx~xc^{(s)}(x) J_0(qx)}_{\displaystyle{c_1}}
- \underbrace{4 \pi \int\limits_0^\infty~dx~e^{-\kappa_Y x} J_0(qx)}_{\displaystyle{c_2}},\nonumber\\~\nonumber
\end{eqnarray}
where $J_0$ denotes the Bessel function of the first kind with index $0$. Note from Fig.~\ref{fig:cx_decay} that the T/2-HNC 
scheme solution for the function $|x c^{(s)}(x)|$ is bound from above by the function $2\Gamma_Y \exp\lbrace -x/2 \rbrace$ for all 
reasonable combinations of $\Gamma_Y$ and $\kappa_Y$, and even in the OCP limit $\kappa_Y = 0$. This finding, combined with the 
upper bound \mbox{$|J_0(x)| < \min \lbrace 1, \sqrt{2/(\pi x)} \rbrace$} for the envelope of the Bessel function, allows us to 
compute an upper bound
\begin{equation}
|c_1| < \min\left\lbrace 12 \pi \Gamma_y, \dfrac{8 \pi \Gamma_Y}{\sqrt{q}} \right\rbrace.
\end{equation}
for the modulus of the function $c_1(q)$. Solutions for all Hankel transforms occurring in such computation are listed in 
Ref.~\cite{Integral1954}. Noting that $c_2 = 4 \pi \Gamma_Y / \sqrt{\kappa_Y^2 + q^2}$, we conclude that $c_1$ in 
Eq.~\eqref{eq:c_qnull_int} is negligible if the condition
\begin{equation}\label{eq:q_kappa_smallness}
q + q_t \gg \kappa_Y^2 + q^2
\end{equation}
is fulfilled, where $q_t \sim 0.1$ is a threshold wavenumber. For all combinations of $q$ and $\kappa_Y$ that fulfill 
Eq.~\eqref{eq:q_kappa_smallness}, the T/2-HNC solution for $c_q$ is well approximated by $c_{q} \approx - c_2 = 4 \pi \Gamma_Y / 
\sqrt{\kappa_Y^2 + q^2}$.


\begin{thebibliography}{42}
\expandafter\ifx\csname natexlab\endcsname\relax\def\natexlab#1{#1}\fi
\expandafter\ifx\csname bibnamefont\endcsname\relax
  \def\bibnamefont#1{#1}\fi
\expandafter\ifx\csname bibfnamefont\endcsname\relax
  \def\bibfnamefont#1{#1}\fi
\expandafter\ifx\csname citenamefont\endcsname\relax
  \def\citenamefont#1{#1}\fi
\expandafter\ifx\csname url\endcsname\relax
  \def\url#1{\texttt{#1}}\fi
\expandafter\ifx\csname urlprefix\endcsname\relax\def\urlprefix{URL }\fi
\providecommand{\bibinfo}[2]{#2}
\providecommand{\eprint}[2][]{\url{#2}}

\bibitem[{\citenamefont{Ivlev et~al.}(2012)\citenamefont{Ivlev, L\"owen,
  Morfill, and Royall}}]{IvlevBook}
\bibinfo{author}{\bibfnamefont{A.}~\bibnamefont{Ivlev}},
  \bibinfo{author}{\bibfnamefont{H.}~\bibnamefont{L\"owen}},
  \bibinfo{author}{\bibfnamefont{G.}~\bibnamefont{Morfill}}, \bibnamefont{and}
  \bibinfo{author}{\bibfnamefont{C.~P.} \bibnamefont{Royall}},
  \emph{\bibinfo{title}{Complex Plasmas and Colloidal Dispersions:
  Particle-Resolved Studies of Classical Liquids and Solids}}
  (\bibinfo{publisher}{World Scientific}, \bibinfo{address}{Singapore},
  \bibinfo{year}{2012}).

\bibitem[{\citenamefont{Pieranski et~al.}(1983)\citenamefont{Pieranski,
  Strzelecki, and Pansu}}]{Pieranski1983}
\bibinfo{author}{\bibfnamefont{P.}~\bibnamefont{Pieranski}},
  \bibinfo{author}{\bibfnamefont{L.}~\bibnamefont{Strzelecki}},
  \bibnamefont{and} \bibinfo{author}{\bibfnamefont{B.}~\bibnamefont{Pansu}},
  \bibinfo{journal}{Phys.~Rev.~Lett.} \textbf{\bibinfo{volume}{50}},
  \bibinfo{pages}{900} (\bibinfo{year}{1983}).

\bibitem[{\citenamefont{Chang and Hone}(1988)}]{Chang1988}
\bibinfo{author}{\bibfnamefont{E.}~\bibnamefont{Chang}} \bibnamefont{and}
  \bibinfo{author}{\bibfnamefont{D.~W.} \bibnamefont{Hone}},
  \bibinfo{journal}{Europhys.~Lett.} \textbf{\bibinfo{volume}{5}},
  \bibinfo{pages}{635} (\bibinfo{year}{1988}).

\bibitem[{\citenamefont{Thomas et~al.}(1994)\citenamefont{Thomas, Morfill,
  Demmel, Goree, Feuerbacher, and M\"ohlmann}}]{Thomas1994}
\bibinfo{author}{\bibfnamefont{H.}~\bibnamefont{Thomas}},
  \bibinfo{author}{\bibfnamefont{G.~E.} \bibnamefont{Morfill}},
  \bibinfo{author}{\bibfnamefont{V.}~\bibnamefont{Demmel}},
  \bibinfo{author}{\bibfnamefont{J.}~\bibnamefont{Goree}},
  \bibinfo{author}{\bibfnamefont{B.}~\bibnamefont{Feuerbacher}},
  \bibnamefont{and}
  \bibinfo{author}{\bibfnamefont{D.}~\bibnamefont{M\"ohlmann}},
  \bibinfo{journal}{Phys. Rev. Lett.} \textbf{\bibinfo{volume}{73}},
  \bibinfo{pages}{652} (\bibinfo{year}{1994}).

\bibitem[{\citenamefont{Fortov et~al.}(2005)\citenamefont{Fortov, Ivlev,
  Khrapak, Khrapak, and Morfill}}]{Fortov2005}
\bibinfo{author}{\bibfnamefont{V.~E.} \bibnamefont{Fortov}},
  \bibinfo{author}{\bibfnamefont{A.~V.} \bibnamefont{Ivlev}},
  \bibinfo{author}{\bibfnamefont{S.~A.} \bibnamefont{Khrapak}},
  \bibinfo{author}{\bibfnamefont{A.~G.} \bibnamefont{Khrapak}},
  \bibnamefont{and} \bibinfo{author}{\bibfnamefont{G.~E.}
  \bibnamefont{Morfill}}, \bibinfo{journal}{Phys. Rep.}
  \textbf{\bibinfo{volume}{421}}, \bibinfo{pages}{1} (\bibinfo{year}{2005}).

\bibitem[{\citenamefont{Fortov and Morfill}(2010)}]{FortovMorfillBook}
\bibinfo{editor}{\bibfnamefont{V.~E.} \bibnamefont{Fortov}} \bibnamefont{and}
  \bibinfo{editor}{\bibfnamefont{G.~E.} \bibnamefont{Morfill}}, eds.,
  \emph{\bibinfo{title}{{Complex and dusty plasmas: from laboratory to space}}}
  (\bibinfo{publisher}{CRC Press, Boca Raton}, \bibinfo{year}{2010}).

\bibitem[{\citenamefont{Konopka et~al.}(2000)\citenamefont{Konopka, Morfill,
  and Ratke}}]{Konopka2000}
\bibinfo{author}{\bibfnamefont{U.}~\bibnamefont{Konopka}},
  \bibinfo{author}{\bibfnamefont{G.~E.} \bibnamefont{Morfill}},
  \bibnamefont{and} \bibinfo{author}{\bibfnamefont{L.}~\bibnamefont{Ratke}},
  \bibinfo{journal}{Phys. Rev. Lett.} \textbf{\bibinfo{volume}{84}},
  \bibinfo{pages}{891} (\bibinfo{year}{2000}).

\bibitem[{\citenamefont{Kompaneets et~al.}(2007)\citenamefont{Kompaneets,
  Konopka, Ivlev, Tsytovich, and Morfill}}]{Kompaneets2007}
\bibinfo{author}{\bibfnamefont{R.}~\bibnamefont{Kompaneets}},
  \bibinfo{author}{\bibfnamefont{U.}~\bibnamefont{Konopka}},
  \bibinfo{author}{\bibfnamefont{A.~V.} \bibnamefont{Ivlev}},
  \bibinfo{author}{\bibfnamefont{V.}~\bibnamefont{Tsytovich}},
  \bibnamefont{and} \bibinfo{author}{\bibfnamefont{G.}~\bibnamefont{Morfill}},
  \bibinfo{journal}{Phys.~Plasmas} \textbf{\bibinfo{volume}{14}},
  \bibinfo{pages}{052108} (\bibinfo{year}{2007}).

\bibitem[{\citenamefont{Vladimirov and Nambu}(1995)}]{Vladimirov1995}
\bibinfo{author}{\bibfnamefont{S.~V.} \bibnamefont{Vladimirov}}
  \bibnamefont{and} \bibinfo{author}{\bibfnamefont{M.}~\bibnamefont{Nambu}},
  \bibinfo{journal}{Phys. Rev. E} \textbf{\bibinfo{volume}{52}},
  \bibinfo{pages}{R2172} (\bibinfo{year}{1995}).

\bibitem[{\citenamefont{Vladimirov and Ishihara}(1996)}]{Vladimirov1996}
\bibinfo{author}{\bibfnamefont{S.~V.} \bibnamefont{Vladimirov}}
  \bibnamefont{and} \bibinfo{author}{\bibfnamefont{O.}~\bibnamefont{Ishihara}},
  \bibinfo{journal}{Phys. Plasmas} \textbf{\bibinfo{volume}{3}},
  \bibinfo{pages}{444} (\bibinfo{year}{1996}).

\bibitem[{\citenamefont{Ishihara and Vladimirov}(1997)}]{Ishihara1997}
\bibinfo{author}{\bibfnamefont{O.}~\bibnamefont{Ishihara}} \bibnamefont{and}
  \bibinfo{author}{\bibfnamefont{S.~V.} \bibnamefont{Vladimirov}},
  \bibinfo{journal}{Phys. Plasmas} \textbf{\bibinfo{volume}{4}},
  \bibinfo{pages}{69} (\bibinfo{year}{1997}).

\bibitem[{\citenamefont{Xie et~al.}(1999)\citenamefont{Xie, He, and
  Huang}}]{Xie1999}
\bibinfo{author}{\bibfnamefont{B.~S.} \bibnamefont{Xie}},
  \bibinfo{author}{\bibfnamefont{K.~F.} \bibnamefont{He}}, \bibnamefont{and}
  \bibinfo{author}{\bibfnamefont{Z.~Q.} \bibnamefont{Huang}},
  \bibinfo{journal}{Phys. Lett. A} \textbf{\bibinfo{volume}{253}},
  \bibinfo{pages}{83} (\bibinfo{year}{1999}).

\bibitem[{\citenamefont{Lemons et~al.}(2000)\citenamefont{Lemons, Murillo,
  Daughton, and Winske}}]{Lemons2000}
\bibinfo{author}{\bibfnamefont{D.~S.} \bibnamefont{Lemons}},
  \bibinfo{author}{\bibfnamefont{M.~S.} \bibnamefont{Murillo}},
  \bibinfo{author}{\bibfnamefont{W.}~\bibnamefont{Daughton}}, \bibnamefont{and}
  \bibinfo{author}{\bibfnamefont{D.}~\bibnamefont{Winske}},
  \bibinfo{journal}{Phys. Plasmas} \textbf{\bibinfo{volume}{7}},
  \bibinfo{pages}{2306} (\bibinfo{year}{2000}).

\bibitem[{\citenamefont{Lapenta}(2000)}]{Lapenta2000}
\bibinfo{author}{\bibfnamefont{G.}~\bibnamefont{Lapenta}},
  \bibinfo{journal}{Phys. Rev. E} \textbf{\bibinfo{volume}{62}},
  \bibinfo{pages}{1175} (\bibinfo{year}{2000}).

\bibitem[{\citenamefont{Schweigert}(2001)}]{Schweigert2001}
\bibinfo{author}{\bibfnamefont{V.~A.} \bibnamefont{Schweigert}},
  \bibinfo{journal}{Plasma Phys. Rep.} \textbf{\bibinfo{volume}{27}},
  \bibinfo{pages}{997} (\bibinfo{year}{2001}).

\bibitem[{\citenamefont{Kompaneets et~al.}(2008)\citenamefont{Kompaneets,
  Vladimirov, Ivlev, and Morfill}}]{Kompaneets2008}
\bibinfo{author}{\bibfnamefont{R.}~\bibnamefont{Kompaneets}},
  \bibinfo{author}{\bibfnamefont{S.~V.} \bibnamefont{Vladimirov}},
  \bibinfo{author}{\bibfnamefont{A.~V.} \bibnamefont{Ivlev}}, \bibnamefont{and}
  \bibinfo{author}{\bibfnamefont{G.~E.} \bibnamefont{Morfill}},
  \bibinfo{journal}{New J. Phys.} \textbf{\bibinfo{volume}{10}},
  \bibinfo{pages}{063018} (\bibinfo{year}{2008}).

\bibitem[{\citenamefont{R\"{o}cker et~al.}(2012)\citenamefont{R\"{o}cker,
  Ivlev, Kompaneets, and Morfill}}]{Roeker2012}
\bibinfo{author}{\bibfnamefont{T.~B.} \bibnamefont{R\"{o}cker}},
  \bibinfo{author}{\bibfnamefont{A.~V.} \bibnamefont{Ivlev}},
  \bibinfo{author}{\bibfnamefont{R.}~\bibnamefont{Kompaneets}},
  \bibnamefont{and} \bibinfo{author}{\bibfnamefont{G.~E.}
  \bibnamefont{Morfill}}, \bibinfo{journal}{Phys. Plasmas}
  \textbf{\bibinfo{volume}{19}}, \bibinfo{pages}{033708}
  (\bibinfo{year}{2012}).

\bibitem[{\citenamefont{Khrapak et~al.}(2010)\citenamefont{Khrapak, Ivlev, and
  Morfill}}]{Khrapak2010}
\bibinfo{author}{\bibfnamefont{S.~A.} \bibnamefont{Khrapak}},
  \bibinfo{author}{\bibfnamefont{A.~V.} \bibnamefont{Ivlev}}, \bibnamefont{and}
  \bibinfo{author}{\bibfnamefont{G.~E.} \bibnamefont{Morfill}},
  \bibinfo{journal}{Phys.~Plasmas} \textbf{\bibinfo{volume}{17}},
  \bibinfo{pages}{042107} (\bibinfo{year}{2010}).

\bibitem[{\citenamefont{Kennedy and Allen}(2003)}]{Kennedy2003}
\bibinfo{author}{\bibfnamefont{R.~V.} \bibnamefont{Kennedy}} \bibnamefont{and}
  \bibinfo{author}{\bibfnamefont{J.~E.} \bibnamefont{Allen}},
  \bibinfo{journal}{J. Plasma Phys.} \textbf{\bibinfo{volume}{69}},
  \bibinfo{pages}{485} (\bibinfo{year}{2003}).

\bibitem[{\citenamefont{Morfill and Ivlev}(2009)}]{Morfill2009}
\bibinfo{author}{\bibfnamefont{G.~E.} \bibnamefont{Morfill}} \bibnamefont{and}
  \bibinfo{author}{\bibfnamefont{A.~V.} \bibnamefont{Ivlev}},
  \bibinfo{journal}{Rev. Mod. Phys.} \textbf{\bibinfo{volume}{81}},
  \bibinfo{pages}{1353} (\bibinfo{year}{2009}).

\bibitem[{\citenamefont{L\"owen}(1992)}]{Lowen92}
\bibinfo{author}{\bibfnamefont{H.}~\bibnamefont{L\"owen}}, \bibinfo{journal}{J.
  Phys.: Condensed Matter} \textbf{\bibinfo{volume}{4}}, \bibinfo{pages}{10105}
  (\bibinfo{year}{1992}).

\bibitem[{\citenamefont{Barreira~Fontecha
  et~al.}(2005)\citenamefont{Barreira~Fontecha, Sch\"ope, K\"onig, Palberg,
  Messina, and L\"owen}}]{Palberg05}
\bibinfo{author}{\bibfnamefont{A.}~\bibnamefont{Barreira~Fontecha}},
  \bibinfo{author}{\bibfnamefont{H.~J.} \bibnamefont{Sch\"ope}},
  \bibinfo{author}{\bibfnamefont{H.}~\bibnamefont{K\"onig}},
  \bibinfo{author}{\bibfnamefont{T.}~\bibnamefont{Palberg}},
  \bibinfo{author}{\bibfnamefont{R.}~\bibnamefont{Messina}}, \bibnamefont{and}
  \bibinfo{author}{\bibfnamefont{H.}~\bibnamefont{L\"owen}},
  \bibinfo{journal}{J. Phys.: Condens. Matter} \textbf{\bibinfo{volume}{17}},
  \bibinfo{pages}{S2779} (\bibinfo{year}{2005}).

\bibitem[{\citenamefont{Abramowitz and Stegun}(1965)}]{Abramowitz}
\bibinfo{author}{\bibfnamefont{L.}~\bibnamefont{Abramowitz}} \bibnamefont{and}
  \bibinfo{author}{\bibfnamefont{I.~A.} \bibnamefont{Stegun}},
  \emph{\bibinfo{title}{Handbook of mathematical functions with formulas,
  graphs, and mathematical tables}} (\bibinfo{publisher}{Dover Publications,
  Inc.}, \bibinfo{address}{New York}, \bibinfo{year}{1965}),
  \bibinfo{edition}{9th} ed., ISBN \bibinfo{isbn}{0-486-61272-4}.

\bibitem[{\citenamefont{Bengtzelius et~al.}(1984)\citenamefont{Bengtzelius,
  G\"otze, and Sj\"olander}}]{Bengtzelius1984}
\bibinfo{author}{\bibfnamefont{U.}~\bibnamefont{Bengtzelius}},
  \bibinfo{author}{\bibfnamefont{W.}~\bibnamefont{G\"otze}}, \bibnamefont{and}
  \bibinfo{author}{\bibfnamefont{A.}~\bibnamefont{Sj\"olander}},
  \bibinfo{journal}{J.~Phys.~C} \textbf{\bibinfo{volume}{17}},
  \bibinfo{pages}{5915} (\bibinfo{year}{1984}).

\bibitem[{\citenamefont{G\"otze}(2009)}]{Goetze2009}
\bibinfo{author}{\bibfnamefont{W.}~\bibnamefont{G\"otze}},
  \emph{\bibinfo{title}{Complex Dynamics of Glass-Forming Liquids: A
  Mode-Coupling Theory}} (\bibinfo{publisher}{Oxford University Press},
  \bibinfo{address}{Oxford}, \bibinfo{year}{2009}).

\bibitem[{\citenamefont{Bayer et~al.}(2007)\citenamefont{Bayer, Brader, Ebert,
  Lange, Fuchs, Maret, Schilling, Sperl, and Wittmer}}]{Bayer2007}
\bibinfo{author}{\bibfnamefont{M.}~\bibnamefont{Bayer}},
  \bibinfo{author}{\bibfnamefont{J.}~\bibnamefont{Brader}},
  \bibinfo{author}{\bibfnamefont{F.}~\bibnamefont{Ebert}},
  \bibinfo{author}{\bibfnamefont{E.}~\bibnamefont{Lange}},
  \bibinfo{author}{\bibfnamefont{M.}~\bibnamefont{Fuchs}},
  \bibinfo{author}{\bibfnamefont{G.}~\bibnamefont{Maret}},
  \bibinfo{author}{\bibfnamefont{R.}~\bibnamefont{Schilling}},
  \bibinfo{author}{\bibfnamefont{M.}~\bibnamefont{Sperl}}, \bibnamefont{and}
  \bibinfo{author}{\bibfnamefont{J.~P.} \bibnamefont{Wittmer}},
  \bibinfo{journal}{Phys.~Rev.~E} \textbf{\bibinfo{volume}{76}},
  \bibinfo{pages}{011508} (\bibinfo{year}{2007}).

\bibitem[{\citenamefont{Fuchs et~al.}(1998)\citenamefont{Fuchs, G\"otze, and
  Mayr}}]{Fuchs1998}
\bibinfo{author}{\bibfnamefont{M.}~\bibnamefont{Fuchs}},
  \bibinfo{author}{\bibfnamefont{W.}~\bibnamefont{G\"otze}}, \bibnamefont{and}
  \bibinfo{author}{\bibfnamefont{M.~R.} \bibnamefont{Mayr}},
  \bibinfo{journal}{Phys.~Rev.~E} \textbf{\bibinfo{volume}{58}},
  \bibinfo{pages}{3384} (\bibinfo{year}{1998}).

\bibitem[{\citenamefont{Franosch et~al.}(1997)\citenamefont{Franosch, Fuchs,
  G\"otze, Mayr, and Singh}}]{Franosch1997}
\bibinfo{author}{\bibfnamefont{T.}~\bibnamefont{Franosch}},
  \bibinfo{author}{\bibfnamefont{M.}~\bibnamefont{Fuchs}},
  \bibinfo{author}{\bibfnamefont{W.}~\bibnamefont{G\"otze}},
  \bibinfo{author}{\bibfnamefont{M.~R.} \bibnamefont{Mayr}}, \bibnamefont{and}
  \bibinfo{author}{\bibfnamefont{A.~P.} \bibnamefont{Singh}},
  \bibinfo{journal}{Phys.~Rev.~E} \textbf{\bibinfo{volume}{55}},
  \bibinfo{pages}{7153} (\bibinfo{year}{1997}).

\bibitem[{\citenamefont{Hajnal et~al.}(2011)\citenamefont{Hajnal, Oettel, and
  Schilling}}]{Hajnal2011}
\bibinfo{author}{\bibfnamefont{D.}~\bibnamefont{Hajnal}},
  \bibinfo{author}{\bibfnamefont{M.}~\bibnamefont{Oettel}}, \bibnamefont{and}
  \bibinfo{author}{\bibfnamefont{R.}~\bibnamefont{Schilling}},
  \bibinfo{journal}{J. Non-Cryst. Solids} \textbf{\bibinfo{volume}{357}},
  \bibinfo{pages}{302} (\bibinfo{year}{2011}).

\bibitem[{\citenamefont{Hansen and McDonald}(1986)}]{Hansen}
\bibinfo{author}{\bibfnamefont{J.-P.} \bibnamefont{Hansen}} \bibnamefont{and}
  \bibinfo{author}{\bibfnamefont{I.~R.} \bibnamefont{McDonald}},
  \emph{\bibinfo{title}{Theory of Simple Liquids}}
  (\bibinfo{publisher}{Academic Press}, \bibinfo{address}{London},
  \bibinfo{year}{1986}), \bibinfo{edition}{3rd} ed.

\bibitem[{\citenamefont{Heinen et~al.}(2014)\citenamefont{Heinen, Allahyarov,
  and L\"{o}wen}}]{Heinen2014}
\bibinfo{author}{\bibfnamefont{M.}~\bibnamefont{Heinen}},
  \bibinfo{author}{\bibfnamefont{E.}~\bibnamefont{Allahyarov}},
  \bibnamefont{and}
  \bibinfo{author}{\bibfnamefont{H.}~\bibnamefont{L\"{o}wen}},
  \bibinfo{journal}{J. Comput. Chem.} \textbf{\bibinfo{volume}{35}},
  \bibinfo{pages}{275} (\bibinfo{year}{2014}).

\bibitem[{\citenamefont{Ng}(1974)}]{Ng1974}
\bibinfo{author}{\bibfnamefont{K.-C.} \bibnamefont{Ng}}, \bibinfo{journal}{J.
  Chem. Phys.} \textbf{\bibinfo{volume}{61}}, \bibinfo{pages}{2680}
  (\bibinfo{year}{1974}).

\bibitem[{\citenamefont{Talman}(1978)}]{Talman1978}
\bibinfo{author}{\bibfnamefont{J.~D.} \bibnamefont{Talman}},
  \bibinfo{journal}{J. Comput. Phys.} \textbf{\bibinfo{volume}{29}},
  \bibinfo{pages}{35} (\bibinfo{year}{1978}).

\bibitem[{\citenamefont{Rossky and Friedman}(1980)}]{Rossky1980}
\bibinfo{author}{\bibfnamefont{P.~J.} \bibnamefont{Rossky}} \bibnamefont{and}
  \bibinfo{author}{\bibfnamefont{H.~L.} \bibnamefont{Friedman}},
  \bibinfo{journal}{J. Chem. Phys.} \textbf{\bibinfo{volume}{72}},
  \bibinfo{pages}{5694} (\bibinfo{year}{1980}).

\bibitem[{\citenamefont{Hamilton}(2000)}]{Hamilton2000}
\bibinfo{author}{\bibfnamefont{A.~J.~S.} \bibnamefont{Hamilton}},
  \bibinfo{journal}{Mon. Not. R. Astron. Soc.} \textbf{\bibinfo{volume}{312}},
  \bibinfo{pages}{257} (\bibinfo{year}{2000}).

\bibitem[{Ham()}]{Hamilton}
\emph{\bibinfo{title}{\mbox{A.~J.~S.~Hamilton's FFTLog website}}},
  \bibinfo{note}{\url{http://casa.colorado.edu/~ajsh/FFTLog/}}.

\bibitem[{\citenamefont{Morita}(1958)}]{Morita1958}
\bibinfo{author}{\bibfnamefont{T.}~\bibnamefont{Morita}},
  \bibinfo{journal}{Prog. Theo. Phys.} \textbf{\bibinfo{volume}{20}},
  \bibinfo{pages}{920} (\bibinfo{year}{1958}).

\bibitem[{\citenamefont{Hartmann et~al.}(2005)\citenamefont{Hartmann, Kalman,
  Donk\'o, and Kutasi}}]{Hartmann2005}
\bibinfo{author}{\bibfnamefont{P.}~\bibnamefont{Hartmann}},
  \bibinfo{author}{\bibfnamefont{G.~J.} \bibnamefont{Kalman}},
  \bibinfo{author}{\bibfnamefont{Z.}~\bibnamefont{Donk\'o}}, \bibnamefont{and}
  \bibinfo{author}{\bibfnamefont{K.}~\bibnamefont{Kutasi}},
  \bibinfo{journal}{Phys. Rev. E} \textbf{\bibinfo{volume}{72}},
  \bibinfo{pages}{026409} (\bibinfo{year}{2005}).

\bibitem[{\citenamefont{Yazdi et~al.}(2014)\citenamefont{Yazdi, Ivlev, Khrapak,
  Thomas, Morfill, L\"owen, Wysocki, and Sperl}}]{Yazdi2014}
\bibinfo{author}{\bibfnamefont{A.}~\bibnamefont{Yazdi}},
  \bibinfo{author}{\bibfnamefont{A.}~\bibnamefont{Ivlev}},
  \bibinfo{author}{\bibfnamefont{S.}~\bibnamefont{Khrapak}},
  \bibinfo{author}{\bibfnamefont{H.}~\bibnamefont{Thomas}},
  \bibinfo{author}{\bibfnamefont{G.~E.} \bibnamefont{Morfill}},
  \bibinfo{author}{\bibfnamefont{H.}~\bibnamefont{L\"owen}},
  \bibinfo{author}{\bibfnamefont{A.}~\bibnamefont{Wysocki}}, \bibnamefont{and}
  \bibinfo{author}{\bibfnamefont{M.}~\bibnamefont{Sperl}},
  \bibinfo{journal}{Phys. Rev. E} \textbf{\bibinfo{volume}{89}},
  \bibinfo{pages}{063105} (\bibinfo{year}{2014}).

\bibitem[{\citenamefont{Baus and Hansen}(1980)}]{Baus1980}
\bibinfo{author}{\bibfnamefont{M.}~\bibnamefont{Baus}} \bibnamefont{and}
  \bibinfo{author}{\bibfnamefont{J.-P.} \bibnamefont{Hansen}},
  \bibinfo{journal}{Phys. Rep.} \textbf{\bibinfo{volume}{59}},
  \bibinfo{pages}{1} (\bibinfo{year}{1980}).

\bibitem[{\citenamefont{Assoud et~al.}(2008)\citenamefont{Assoud, Messina, and
  L\"owen}}]{Messina08}
\bibinfo{author}{\bibfnamefont{L.}~\bibnamefont{Assoud}},
  \bibinfo{author}{\bibfnamefont{R.}~\bibnamefont{Messina}}, \bibnamefont{and}
  \bibinfo{author}{\bibfnamefont{H.}~\bibnamefont{L\"owen}},
  \bibinfo{journal}{J. Chem. Phys.} \textbf{\bibinfo{volume}{129}},
  \bibinfo{pages}{164511} (\bibinfo{year}{2008}).

\bibitem[{\citenamefont{Erd\'{e}lyi}(1954)}]{Integral1954}
\bibinfo{editor}{\bibfnamefont{A.}~\bibnamefont{Erd\'{e}lyi}}, ed.,
  \emph{\bibinfo{title}{Tables of integral transforms. Based, in part, on notes
  left by Harry Bateman Volume II}} (\bibinfo{publisher}{McGraw-Hill},
  \bibinfo{address}{New York}, \bibinfo{year}{1954}).

\end{thebibliography}
\end{document}